\documentclass[twocolumn,10pt,showpacs,preprintnumbers,amsmath,amssymb]{revtex4}

\usepackage{graphicx}
\usepackage{dcolumn}
\usepackage{bm}

\begin{document}

\title{Vibron-polaron critical localization in a finite size molecular nanowire}

\author{C. Falvo and V. Pouthier$^{*}$}

\address{Laboratoire de Physique Mol\'{e}culaire, UMR CNRS 6624. Facult\'{e} des
Sciences - La Bouloie, \\ Universit\'{e} de Franche-Comt\'{e}, 25030 Besan\c {c}on cedex, 
France.}

\email{vincent.pouthier@univ-fcomte.fr}

\date{\today}

\begin{abstract}
The small polaron theory is applied to describe the vibron dynamics in an adsorbed nanowire with a special emphasis onto finite size effects. It is shown that the finite size of the nanowire discriminates between side molecules and core molecules which experience a different dressing mechanism. Moreover, the inhomogeneous behavior of the polaron hopping constant is established and it is shown that the core hopping constant depends on the lattice size. However, the property of a lattice with translational invariance is recovered when the size of the nanowire is greater than a critical value. Finally, it is pointed out that these features yield the occurrence of high energy localized states which both the nature and the number are summarized in a phase diagram in terms of the relevant parameters of the problem (small polaron binding energy, temperature, lattice size).
\end{abstract}


\maketitle

\section{Introduction}

As a result of following Moore's law during the last 40 years \cite{kn:moore}, the industry now manufactures integrated circuits with feature sizes bellow 180 nm. To gain one or two orders of magnitude in these feature sizes constitutes one of the challenges of the modern technology. In that context, among the different ways investigated to control the information transfer at the nanoscale, the use of adsorbed nanostructures appears promising \cite{kn:dash,kn:lagally,kn:bruch,kn:duke}. Indeed, it is now well established that surfaces exhibiting self-organized defects are ideal templates for the formation of low dimensional nanostructures such as 1D wires and confined 2D monolayers \cite{kn:sundaram,kn:kern1,kn:kern2}. In addition, the interest in adsorbed nanodevices is reinforced by the fact that local probes, such as scanning tunneling microscope (STM), can serve as tools to design the nanostructure \cite{kn:meyer,kn:binnig,kn:dujardin,kn:ho1,kn:ho2} and to excite its electronic or vibrational degrees of freedom \cite{kn:ho3,kn:ho4}.

Although the electron transport is expected to play a key role in such nanodevices \cite{kn:pouthierelec1,kn:pouthierelec2}, the ability to excite the vibration of the admolecules with a STM were exploited in recent works to propose an alternative based on the use of vibrational excitons \cite{kn:pouthier1,kn:pouthier2,kn:pouthier3}. Indeed, in a molecular lattice, the lateral interactions induce a coupling between the internal degrees of freedom of the molecules and favor the coherent propagation of the internal vibrations from one site to another. The delocalization of the vibrational energy yields the formation of vibrational excitons called vibrons. 

In a nanowire, the vibrons do not propagate freely but interact with their surrounding formed by the low frequency modes of the adsorbates and by the phonons and the electrons of the substrate.  All these excitations are responsible for both energy and phase relaxation, which the knowledge is essential to understand and exploit the vibronic transport. Note that using the vibrons as a vehicle for the information transfer requires their confinement inside the adsorbate so that metallic surfaces cannot be used. Indeed, fast energy relaxation takes place because of the resonances of the vibrons with the electron-hole pair continuum of the metal which leads to a shortening of the vibron lifetime, typically about 1 ps \cite{kn:lifetime}. 
As a result, we restrict our attention onto the use of dielectric surfaces for which energy relaxation occurs via multiphonon excitations, only, leading to a longer lifetime. For instance, the vibrational lifetime for the Si-H vibration for the system Si-H is about 1 ns at room temperature \cite{kn:guyot1,kn:guyot2}. A remarkable example is given by the stretching vibration of CO adsorbed on NaCl for which the lifetime was measured to be about 0.01 s \cite{kn:ewing}.

In that context, most of the theoretical analysis of the vibron dynamics were applied to  interpret the infrared spectra with a special emphasis on the processes responsible for line broadening \cite{kn:tobin,kn:gadzuk,kn:persson1,kn:persson2,kn:persson3,kn:langreth,kn:rosembaum1,kn:rosembaum2,kn:pouthier4,kn:pouthier5}. Dephasing mechanisms were thus accounted by 
assuming a sufficiently weak vibron-phonon coupling and by invoking the Markovian limit. Within the standard perturbative theory, the effect of the bath was summarized in a simple parameter : the so-called dephasing constant.

In the present paper, we point out that these assumptions are questionable to treat the vibron dynamics in a nanoscale molecular adsorbate. Indeed, it is well known that such systems exhibit rather small vibron bandwidths, typically about a few wave numbers \cite{kn:guyot1,kn:pouthier4}. Therefore, it seems reasonable to assume that the vibron hopping constant is about or less than the strength of the vibron-phonon coupling so that the standard perturbative theory cannot be applied. In addition, the time spends by a vibron to propagate coherently along one nanometer is about a few picoseconds, i.e. the same order of magnitude than the admitted value for the thermal bath correlation time. As a consequence, non Markovian effects are expected to play a key role for nanoscale transport.

To overcome these difficulties, we introduce a small polaron theory to describe the vibron transport. In analogy with the vibrational dynamics in model proteins \cite{kn:davydov}, the vibron dynamics is supposed to be governed by the so-called dressing effect \cite{kn:brown1,kn:brown2,kn:ivic1,kn:ivic2,kn:ivic3,kn:ivic4,kn:pouthier6,kn:pouthier7}.
When the vibron-phonon interaction is sufficiently strong, the creation of a vibron is accompanied by a lattice distortion. However, since the vibron bandwidth is smaller than the phonon cutoff frequency, the non-adiabatic limit is reached \cite{kn:ivic2}. The lattice distortion follows instantaneously the vibron during its propagation so that the vibron is dressed by a virtual cloud of phonons and forms the small polaron. Although the dressing effect is well characterized in a lattice with translational invariance, very little is known about finite size effects. Therefore, in the present paper, the small polaron theory is applied to an adsorbed nanowire with a special emphasis onto the symmetry breaking induced by the vibron-phonon confinement. Note that the present paper is devoted to the characterization of the polaron eigenstates, only. The study of the vibron dynamics versus time will be addressed in a forthcoming work. 

The paper is organized as follows. In Sec. II, the vibron-phonon Hamiltonian used to described the finite size nanowire is introduced. In Sec. III, the nature of the phonon states in a confined nanowire is first summarized. Then, a Lang-Firsov \cite{kn:lang} transformation is applied to renormalize the vibron-phonon interaction and to reach the small polaron point of view. Finally, a mean field procedure is performed to obtain the effective Hamiltonian for the dressed vibrons. In Sec. IV, a detailed analysis of the polaron eigenstates is performed and the results are discussed and interpreted in Sec. V.

\section{Model and Hamiltonian}

Let us consider a set of $N$ molecules adsorbed on the surface of a well-organized substrate.
The molecules form a finite size 1D lattice which the site position is denoted $n=1,2,...,N$. 

To characterize the vibron dynamics, we assume that each molecule $n$ behaves as an internal high frequency oscillator described by the standard vibron operators $b^{+}_{n}$ and $b_{n}$. The vibron Hamiltonian is thus expressed as (using the convention $\hbar=1$)
\begin{equation}
H_{v}=\sum_{n=1}^{N} \omega_{0}b^{\dag}_{n}b_{n}+\sum_{n=1}^{N-1}J [b^{\dag}_{n}b_{n+1}+b^{\dag}_{n+1}b_{n}]
\label{eq:Hv}
\end{equation}
where $\omega_{0}$ stands for the internal frequency of each molecule and where $J$ denotes the vibron hopping constant between nearest neighbor admolecules.
Note that long range lateral interactions may affect the vibron dynamics but these effects are expected to be rather weak in a 1D lattice. 

To mimic the influence of the phonon bath in a simple way, we assume that the internal vibration of each molecule interacts preferentially with the longitudinal phonons of the nanowire. These phonons refer to the collective dynamics of the center of mass of the admolecules around their equilibrium positions. Within the harmonic approximation, the phonon Hamiltonian is thus written as
\begin{equation}
H_{p}=\sum_{n=1}^{N} \frac{p_{n}^{2}}{2M}+\sum_{n=1}^{N-1}\frac{1}{2}W(u_{n+1}-u_{n})^{2}
\label{eq:Hp}
\end{equation}
where $u_{n}$ and $p_{n}$ denote the displacement and the corresponding momentum of the $n$ th admolecule, respectively. $M$ stands for the mass of each admolecule which interact with its nearest neighbor molecules via the lateral force constant $W$.

By neglecting the population relaxation, the phonon bath leads to stochastic fluctuations of the dynamical parameters which characterize the vibron dynamics. In this work, we use the potential deformation model \cite{kn:dibartolo,kn:skinner,kn:berne,kn:golosov} which supposes that the bath induces a random modulation of the internal frequency of each molecule. Within this model, the vibron-phonon coupling Hamiltonian is expressed as
\begin{eqnarray}
\Delta H_{vp}&=&\sum_{n=2}^{N-1} \chi(u_{n+1}-u_{n-1})b^{\dag}_{n}b_{n} \nonumber \\
&+&\chi(u_{2}-u_{1})b^{\dag}_{1}b_{1}+\chi(u_{N}-u_{N-1})b^{\dag}_{N}b_{N}
\label{eq:Hvp}
\end{eqnarray}
where $\chi$ is the strength of the vibron-phonon coupling. Note that the previous model cannot be applied to molecules on metals. Indeed, Persson and co-workers (see for instance \cite{kn:persson1,kn:persson2,kn:persson3}) have shown that the high frequency stretching mode is preferentially coupled to either the frustrated translational motion or the frustrated rotational motion of the admolecule. Since the admolecules are adsorbed onto high symmetry adsorption sites, the vibron-phonon coupling depends in a quadratic way on the phonon coordinates. However, as pointed out in the introduction, we restrict our attention to admolecules weakly bounded to the surface for which the preferential pathway for dephasing involves the collective modes of the monolayer. This situation occurs, for instance, for the CO/NaCl system \cite{kn:pouthier4}.   

Finally, the vibron-phonon dynamics in the nanowire is governed by the full Hamiltonian $H=H_{v}+H_{p}+\Delta H_{vp}$. The following section is devoted to its simplification and to the introduction of an effective Hamiltonian for the vibrons, only.

\section{Theoretical Background}

As mentioned in the introduction, the vibron dynamics in a molecular lattice with a strong  vibron-phonon interaction is essentially governed by the dressing effect. The characterization of this effect requires the knowledge of the phonon eigenstates and of the Lang-Firsov transformation \cite{kn:lang} which yields the polaron point of view. However, although these two features are well known in an infinite lattice with translational invariance, finite size induces a symmetry breaking responsible for a modification of both the phonon structure and the dressing mechanism. The present section is thus devoted to the presentation of these modifications. 

\subsection{phonon states}

In a finite size nanowire, the phonon states do not correspond to Bloch waves with well defined wave vectors. Indeed, the sides of the lattice lead to a reflexion of the acoustic waves 
so that the true eigenstates appear as a superimposition of incident and reflected plane waves.
More precisely, the nanowire, which is responsible for the confinement of the phonons, 
behaves as a resonant cavity. It selects particular values of the wave vector due to the free boundary conditions so that a stationary regime takes place. 

Therefore, it is straightforward to show that the phonon eigenstates correspond to $N$ normal modes with quantized wave vectors $q_{p}=p\pi/N$, $p=0,1,..,N-1$ and frequency $\Omega_{p}=\Omega_{c} \sin(p\pi/2N)$, where $\Omega_{c}=\sqrt{4W/M}$ denotes the cutoff frequency of the corresponding infinite nanowire. 
Within this normal mode representation, the phonon Hamiltonian $H_{p}$ (Eq.(\ref{eq:Hp})) is thus rewritten as 
\begin{equation}
H_{p}=\sum_{p=0}^{N-1} \Omega_{p}(a_{p}^{\dag}a_{p}+1/2)
\label{eq:Hp1}
\end{equation}
The vibron-phonon coupling Hamiltonian $\Delta H_{vp}$ (Eq.(\ref{eq:Hvp})) 
is expressed as
\begin{equation}
\Delta H_{vp}=\sum_{n=1}^{N} \sum_{p=1}^{N-1} \chi_{p,n}(a_{p}^{\dag}+a_{p})b^{\dag}_{n}b_{n}
\label{eq:Hvp1}
\end{equation}
where $\chi_{p,n}$ characterizes the strength of the coupling between the vibrons and the $p$ th normal mode. From Eqs.(\ref{eq:Hvp}), it is straightforward to show that 
\begin{equation}
\chi_{p,n}=-\sqrt{\frac{2}{N}}\Delta_{0}\frac{\sin(p\pi/N)}{\sqrt{\mid \sin(p\pi/2N) \mid}} \sin(\frac{p \pi}{N}(n-1/2))
\label{eq:chi}
\end{equation}
where $\Delta_{0}=\chi (\hbar^{2} M W)^{-1/4}$ ($\hbar$ as be reintroduced to avoid confusion).
Note that the mode with zero wave vector describes the free translation of the nanowire. 
However, since this mode disappears in an adsorbed nanowire due to its interaction with the substrate, it will be disregarded in the following of the text.

\subsection{Small polaron theory and effective Hamiltonian}

To partially remove the vibron-phonon coupling Hamiltonian, a Lang-Firsov transformation is applied \cite{kn:lang}. Indeed, since the vibron-phonon dynamics is dominated by the so-called dressing effect, we consider a "full dressing" and introduce the following unitary transformation
\begin{equation}
U=\exp(\sum_{n=1}^{N}\sum_{p=1}^{N-1}\frac{\chi_{n,p}}{\Omega_p} (a_p^\dag-a_p)b_n^\dag b_n)
\label{eq:U}
\end{equation}
By using Eq.(\ref{eq:U}), the transformed Hamiltonian $\hat{H}=U H U^\dag$
is written as
\begin{eqnarray}
&&\hat{H}= \sum_{n=1}^{N} (\omega_{0}-\epsilon_{n})b^\dag_n b_n-\epsilon_{n} b^{\dag 2}_nb_n^2 \nonumber \\ &+&\sum_{n=1}^{N-1} J[\Theta_{n}^\dag\Theta_{n+1}b^\dag_{n}b_{n+1}+h.o.]-E_Bb^\dag_{n+1}b^\dag_{n}b_{n+1}b_{n} \nonumber \\ 
&+&\sum_{p=1}^{N-1} \Omega_pa^\dag_pa_p
\label{eq:HHAT}
\end{eqnarray}
where h.o. denotes the hermitian operator and 
where $\epsilon_{n}= E_{B}(1-1/2(\delta_{n,1}+\delta_{n,N}))$ is expressed in terms of the small polaron binding energy $E_{B}=2\Delta_{0}^{2}/\Omega_{c}$. In Eq.(\ref{eq:HHAT}),  $\Theta_{n}$ stands for the dressing operator defined as
\begin{equation}
\Theta_{n}=\exp(-\sum_{p=1}^{N-1}\frac{\chi_{n,p}}{\Omega_p} (a_p^\dag-a_p))
\label{eq:theta}
\end{equation}

In this dressed vibron point of view (Eq.(\ref{eq:HHAT})), the vibron-phonon coupling remains through the modulation of the lateral terms by the dressing operators. Although these operators depend on the phonon coordinates in a highly nonlinear way, the vibron-phonon interaction has been strongly reduced within this transformation. As a result, we can take advantage of such a reduction to perform a mean field procedure \cite{kn:ivic1,kn:ivic2} and to express the full Hamiltonian $\hat{H}$ as the sum of three separated contributions as 
\begin{equation}
\hat{H}=\hat{H}_{eff}+H_{p}+\Delta H 
\end{equation}
where $\hat{H}_{eff}=\langle(\hat{H}-H_{p})\rangle$ denotes the effective Hamiltonian of the dressed vibrons and where $\Delta H =\hat{H}-H_{p}-\langle (\hat{H}-H_{p})\rangle$ stands for the remaining part of the vibron-phonon interaction. The symbol $\langle\hdots\rangle$ represents a thermal average over the phonon degrees of freedom which are assumed to be in thermal equilibrium at temperature $T$.

As a result, the effective dressed vibron Hamiltonian is written as
\begin{eqnarray}
&&\hat{H}_{eff}=\sum_{n=1}^N (\omega_{0}-\epsilon_{n})b^\dag_nb_n-
\epsilon_{n}b^{\dag2}_nb_n^2 \nonumber \\ 
&+&\sum_{n=1}^{N-1}\hat{J}_{n}[b^\dag_{n}b_{n+1}+h.o.]-E_Bb^\dag_{n+1}b^\dag_{n}b_{n+1}b_{n} 
\label{eq:HHATEFF}
\end{eqnarray}
where $\hat{J}_{n}=J\exp(-S(T,n))$ and where $S(T,n)$ is the coupling constant defined as 
($k_{B}$ denotes the Boltzmann constant)
\begin{equation}
S(T,n)=\frac{1}{2}\sum_{p=1}^{N-1} ( \frac{\chi_{n,p}-\chi_{n+1,p}}{\Omega_p})^2 \coth(\frac{\Omega_p}{2k_BT})
\label{eq:S}
\end{equation}

The Hamiltonian $\hat{H}_{eff}$ (Eq.(\ref{eq:HHATEFF})) describes the dynamics of vibrons dressed by a virtual cloud of phonons, i.e. small polarons. It accounts for a renormalization of the main part of the vibron-phonon coupling within the non-adiabatic limit. The remaining vibron-phonon coupling is thus assumed to be small in order to be treated using perturbative theory. Such a contribution, disregarded in the present work, will be addressed in a forthcoming paper. 
 
Eq.(\ref{eq:HHATEFF}) clearly shows the interplay between the usual dressing effect and the finite size of the nanowire. 
Indeed, within our procedure, the main features induced by the dressing effect in an infinite lattice are recovered. More precisely, the dressing modifies the harmonic part of the Hamiltonian and yields a redshift of each internal frequency. Moreover, it reduces the strength of the vibron hopping constant so that the effective mass of the polarons increases. In addition, the dressing favors the occurrence of coupling terms which break the vibron independence and which directly affect the multi-vibron dynamics \cite{kn:pouthier6,kn:pouthier7}. These terms will be disregarded in the present work devoted to the single-vibron dynamics.  

However, the finite size of the lattice discriminates between the core molecules, i.e. those which are located at the sites $n=2,...,N-1$, and the side molecules $n=1$ and $n=N$. Indeed, Eq.(\ref{eq:Hvp}) shows that core and side molecules do not experience the same interaction with the phonon bath which results in a different dressing effect. The redshift of the frequency of the core molecules, equal to the small polaron binding energy $E_{B}$, corresponds to the shift occurring in an infinite lattice. In a marked contrast, the side molecules experience a frequency shift two times smaller. In other words, for the polaron dynamics, the finite size nanowire exhibits two defects located at the side sites $n=1$ and $n=N$. 
In addition, the modification of the nature of the phonons induced by their confinement favors an inhomogeneity in the polaron hopping constant. Indeed, in a marked contrast with the translational invariant lattice, the coupling constant $S(T,n)$ depends on the site position. As a result, the effective hopping constant $\hat{J}_{n}=J\exp(-S(T,n))$ is not uniform along the nanowire and enhances the symmetry breaking induced by the confinement. 

At this step, the restriction of the effective Hamiltonian $\hat{H}_{eff}$ (Eq.(\ref{eq:HHATEFF})) to the single-vibron subspace can be easily diagonalized by using standard numerical procedures. As a result, the behavior of the single-vibron eigenstates can be characterized depending on the values of the parameters involved in the model as illustrated in the following section.

\section{Numerical results}

In this section, the previous formalism is applied to characterize the small polaron eigenstates in a finite size nanowire. The undressed hopping constant is fixed to the typical value $J=4.0$ cm$^{-1}$ whereas the phonon cutoff frequency is set to $\Omega_{c}=100$ cm$^{-1}$. Note that the frequencies are centered around the vibrational frequency $\omega_{0}$ set to zero. The temperature $T$, the small polaron binding energy $E_{B}$ and the lattice size $N$ are parameters allowed to vary.

In Fig. 1, the dependence of the polaron hopping constant $\hat{J}_{n}$ on the position of the sites is drawn for $T=5$ K (Fig. 1a) and $T=50$ K (Fig. 1b). The number of molecules is fixed to $N=20$ and three typical values for $E_{B}$ are considered. Whatever the temperature, the hopping constant decreases as $E_{B}$ increases as a result of the well-known dressing effect. 
At low temperature (Fig. 1a), the hopping constant displays a rather important inhomogeneity which increases as $E_{B}$ increases. Although $\hat{J}_{n}$ is almost uniform in the core of the nanowire, it increases close to the lattice sides where a reduced dressing effect takes place. For weak $E_{B}$ values, i.e. $E_{B}=10$ cm$^{-1}$, only the hopping constant involving side molecules differs significantly from the hopping constant in the core of the nanowire. For instance, the ratio $\hat{J}_{1}/J=0.95$ is slightly greater than the core ratio equal to $\hat{J}_{N/2}/J=0.92$. By contrast, as when increasing $E_{B}$, both the strength and the range of the inhomogeneity increase. When $E_{B}=100$ cm$^{-1}$, the ratio $\hat{J}_{1}/J=0.58$ is 1.3 times greater than the core ratio equal to $\hat{J}_{N/2}/J=0.43$ and the hopping constant $\hat{J}_{2}/J=0.45$ slightly differs form the core value. At high temperature (Fig. 1b), the inhomogeneous effects are drastically reduced. Only the hopping constant involving side molecules differs form its core value whatever $E_{B}$. However, this difference represents less than 10$\%$ of the core hopping constant. Note that as when increasing again the temperature, the inhomogeneous effects almost disappear. 

To characterize the inhomogeneous nature of the hopping constant, Fig. 2 represents the behavior of the difference between the side and the core hopping constants with respect to $E_{B}$ for $N=30$ and $T=5$ K (full circles), $50$ K (open circles) and $100$ K (full triangles). The illustrated features coroborate the previous results since it is shown that the difference between the side and the core hopping constants increases as $E_{B}$ increases and as the temperature decreases.

Finally, the behavior of the core hopping constant with respect to the size $N$ of the nanowire is illustrated in Fig. 3 for $T=50$ K and for $E_{B}=10$ cm$^{-1}$ (full circles), $50$ cm$^{-1}$ (open circles) and $100$ cm$^{-1}$ (full triangles). As previously, the decrease of the core hopping constant with the small polaron binding energy is a signature of the dressing effect. However, Fig. 3 clearly shows that the dressing mechanism is size dependent and the core hopping constant exhibits two regimes with respect to $N$. For small sizes, typically when $N<20$, the core hopping constant decreases rapidly as $N$ increases. Note that it is equal to the undressed value $J$ when $N=2$. By contrast, when the lattice size is sufficiently important, the core hopping constant converges to a constant value equal to the polaron hopping constant in an infinite lattice. In other words, the dressing effect decreases as the size of the nanowire decreases.

In Figs. 4, 5 and 6, the polaron eigenstates of a nanowire with $N=15$ molecules are characterized for three different $E_{B}$ values. The temperature is fixed to $T=100$ K so that the inhomogeneity of the polaron hopping constant affects the sides of the nanowire, only. We thus consider that the hopping constant is uniform in the core of the lattice and equal to its value $\hat{J}_{N/2}$ at the center of the nanowire. Such a behavior allows us to introduce the polaron band which the eigenenergies $\omega$ range between $\omega_{0}-E_{B}-2\hat{J}_{N/2}$ and $\omega_{0}-E_{B}+2\hat{J}_{N/2}$. This band contains the single polaron states of an infinite nanowire, i.e. Bloch waves, characterized by the finite size hopping constant $\hat{J}_{N/2}$.
Therefore, to characterize this band for the three different $E_{B}$ values, the corresponding densities of states are drawn (full line) in Figs. 4a, 5a and 6a. 

When $E_{B}=3$ cm$^{-1}$ (Fig. 4a), the core hopping constant is equal to $\hat{J}_{N/2}=3.71$ cm$^{-1}$ and the polaron band extends from $-10.42$ cm$^{-1}$ to $4.42$ cm$^{-1}$ (around $\omega_{0}$). The finite size nanowire exhibits $N=15$ eigenstates which the energies are located inside the band. To illustrate the nature of these states, the wave function of the two states with the two higher energies are drawn in Fig. 4b. The figure clearly shows that these states correspond to the first two stationary states of a finite size lattice. In fact, in that case, our calculations establish that the $N$ eigenstates are $N$ stationary states. 

When $E_{B}=15$ cm$^{-1}$ (Fig. 5a), the polaron bandwidth is reduced ($\hat{J}_{N/2}=2.74$ cm$^{-1}$) and extends from $-20.48$ cm$^{-1}$ to $-9.52$ cm$^{-1}$. The nanowire supports $N-2=13$ states which the energies are located inside the band and exhibits two states outside the polaron band. The corresponding energies, almost degenerated and equal to $-6.49$ cm$^{-1}$, are greater than the top of the band. As shown in Fig. 5b, these two states correspond to two localized states which the wave function is important close to the sides of the nanowire. Note that one state appears symmetrically localized onto each side whereas the other one exhibits an anti-symmetric localization. 

As shown in Fig. 6a, an intermediate situation occurs when $E_{B}=7.2$ cm$^{-1}$. In that case, the polaron band extends from $-13.86$ cm$^{-1}$ to $-0.54$ cm$^{-1}$ which corresponds to a core hopping constant $\hat{J}_{N/2}=3.33$ cm$^{-1}$. Our calculations clearly show that a single eigenstate has an energy equal to $-0.48$ cm$^{-1}$ greater than the top of the band. A shown in Fig. 6b, this state corresponds to a state symmetrically localized onto each side. However, the localization length is larger than or about to the system size so that the wave function exhibits a significant value over the entire lattice.  

The previous results illustrate the fact that the confined nanowire exhibits localized states depending on the values of the small polaron binding energy. In fact, our calculations have established that the finite size lattice supports at least zero, one or two localized states.   
To characterize this feature, the behavior of the energy of the localized states vs $E_{B}$ is displayed in Fig. 7 for different $N$ values and for $T=100$ K. Note that the origin of the energy  is taken at the top of the polaron band.
The figure clearly shows that the number of localized states depends on the value of both $E_{B}$ and $N$. 
When $N=10$, the first localized state occurs when $E_{B}=6.79$ cm$^{-1}$ (full circles) whereas the second localized state takes place when $E_{B}=8.06$ cm$^{-1}$ (full triangle). The energy of each localized state scales similarly with respect to the small polaron binding energy. More precisely, the energy exhibits a power law dependence close to the top of the band and reaches a linear regime for sufficiently strong $E_{B}$ values. In this linear regime, the energies of the two localized states are almost degenerated. 
When $N=15$, the first localized state occurs when $E_{B}=6.71$ cm$^{-1}$ (open circle), i.e. almost the same $E_{B}$ value responsible for the occurrence of the first localized state when $N=10$. By contrast, the increase of the size of the nanowire reduces the critical $E_{B}$ value for the occurrence of the second localized states. Indeed, when $N=15$, this latter state takes place when $E_{B}=7.51$ cm$^{-1}$ (open triangles). 

In the same way, Fig. 8 shows the behavior of the energy of the localized states versus the temperature for different $N$ values and for $E_{B}=7$ cm$^{-1}$. The figure clearly shows that the nanowire supports zero localized states at low temperature. However, as when increasing the temperature, two transitions appear successively leading to the occurrence of two localized states. When $N=10$, the first localized state occurs at $T=50 $ K (full circles) whereas the second localized state takes place at $T=146$ K (full triangle). The energy of each localized state scales similarly with respect to the temperature. It first exhibits a power law dependence close to the top of the polaron band and reaches a linear regime at a sufficiently high temperature. In the linear regime, the energies of the two localized states are degenerated. 
When $N=15$, the first localized state occurs at $T=45$ cm$^{-1}$ (open circles) whereas the second localized state is created at $T=106$ K (open triangles). Note that as when increasing the value of the small polaron energy, the critical temperatures at which the localization takes place are reduced.

\section{Discussion}

To interpret and discuss the previous numerical results, let us first focus our attention onto the nature of the polaron hopping constant. Indeed, it has been shown (see Figs. 1-3) that the finite size of the nanowire favors an inhomogeneous behavior of the hopping constant which appears site dependent. However, such a feature occurs essentially at low temperature and strong small polaron binding energy and the hopping constant becomes almost site independent at high temperature. 

Indeed, in this latter situation, it is straightforward to show that the coupling constant $S(T,n)$ Eq.(\ref{eq:S}) is expressed as 
\begin{equation}
S(T,n)=\frac{4E_B k_B T}{\Omega_{c}^2}(1-\frac{2}{N})
\label{eq:SHT}
\end{equation}
As a consequence, the effective hopping constant within the high temperature limit is written as
\begin{equation}
\hat{J}_{n} = J(\frac{\hat{J}_{\infty}}{J})^{1-2/N}=\hat{J}_{\infty}e^{\frac{N^{*}}{N}}
\label{eq:Jeff}
\end{equation}
where $\hat{J}_{\infty}$ denotes the polaron hopping constant in an infinite nanowire and where $N^{*}\propto E_{B}k_{B}T/\Omega_{c}^{2}$. Eq.(\ref{eq:Jeff}) shows that the high temperature polaron hopping constant is translationally invariant but depends on the size of the nanowire. More precisely, it increases as the lattice size decreases to reach its undressed value equal to $J$ when the nanowire reduces to a dimer with $N=2$. In that case, the phonon bath induces the same frequency modulation for the two molecules $n=1$ and $n=2$ (see Eq.(\ref{eq:Hvp}) when $N=2$). Therefore, the resonance between the two internal modes is never broken by the phonon motion so that nothing prevents the delocalization of the vibrational energy. The vibrational energy flow occurs according to the rate $J$ which characterizes undressed molecules. By contrast, as when increasing the lattice size, the polaron hopping reach the limiting value $\hat{J}_{\infty}$ of the effective hopping constant in an infinite lattice.  

In other words, the dressing effect is size dependent. This feature originates in the modification of the phonon states induced by their confinement in the finite size nanowire. Indeed, this confinement is responsible for the occurrence of a finite number of stationary states characterized by a discrete energy spectrum. As decreasing the lattice size, the number of phonon modes decreases so that the vibron-phonon interaction is reduced when compared to that in a lattice with the translational invariance. However, our results clearly show that the property of the polaron band is recovered when the size of the nanowire is greater than the typical value $N^{*}$ (see Fig. 3). 

At low temperature, no analytical expression of the coupling constant can be obtained. However, we can take advantage of the fact that the dependence of the hopping constant on the site position is important in the vicinity of the nanowire sides, only (see Fig.1). Therefore, to described such a behavior, we consider a semi-infinite nanowire and perform the calculation of Eq.(\ref{eq:S}) within the limit $N \rightarrow \infty$. As a result, at zero temperature, the coupling constant near the side $n=1$ is expressed as
\begin{equation}
S(0,n) = \frac{16 E_B}{3 \pi \Omega_c} \frac{9-104 n^2+128 n^4}{9-160 n^2 +256 n^4}
\label{eq:SLT}
\end{equation}
Eq.(\ref{eq:SLT}) allows us to reproduce the behavior of the polaron hopping constant in the vicinity of the side $n=1$. It shows that the range of the inhomogeneity of the hopping constant depends on the ratio $E_{B}/\Omega_{c}$. In other words when $E_{B}$ is smaller than the phonon cutoff frequency the inhomogeneous effects disappear as shown in Fig. 1. 

The previous features allow us to characterize the single-polaron dynamics in a simple way in order to understand the mechanisms involved in the occurrence of localized states. Indeed, the numerical results clearly show that the inhomogeneity of the hopping constant affects the sides of the nanowire, only. Therefore, we suppose that the hopping constant is uniform in the core of the lattice and equal to its value $\hat{J}_{N/2}$ at the center of the nanowire. By contrast, we assume that hopping processes involving side molecules is singular and characterized by the constant  $\hat{J}_{1}=\hat{J}_{N-1}$ so that the hopping constant is approximated as
\begin{eqnarray}
\hat{J}_{n} = \left\{ \begin{array}{cc} 
\hat{J}_{N/2} & \mbox{if $n=2,...,N-2$} \\
\hat{J}_{1} & \mbox{if $n=1$ and $n=N-1$}
\end{array}
\right.
\label{eq:JHAT}
\end{eqnarray}
As a result, by inserting Eq.(\ref{eq:JHAT}) into the expression of the effective polaron Hamiltonian $\hat{H}_{eff}$ Eq.(\ref{eq:HHATEFF}) yields the Schrodinger equation $\hat{H}_{eff}\mid \psi \rangle = \omega \mid \psi \rangle$ for single-polaron states as
\begin{eqnarray}
\lambda \psi_1 &=& \alpha \psi_1+ (1+\beta) \psi_{2} \nonumber \\
\lambda \psi_2 &=& (1+\beta)\psi_{1} +\psi_{3} \nonumber \\
\lambda \psi_3 &=& \psi_{2} +\psi_{4} \nonumber \\
...&=&... \nonumber \\
\lambda \psi_N &=& \alpha \psi_N+ (1+\beta) \psi_{N-1} 
\label{eq:schrod2}
\end{eqnarray}
where $\psi_n$ denotes the polaron wave function onto the $n$ th site and where $\lambda=(\omega-\omega_{0}+E_{B})/\hat{J}_{N/2}$, $\alpha=E_{B}/2\hat{J}_{N/2}$ and $\beta=(\hat{J}_{1}-\hat{J}_{N/2})/\hat{J}_{N/2}$. 

From a physical point of view, the wave function $\psi_{n}$ is a superimposition of Bloch waves $\psi_{n}=A^{+}e^{ikn}+A^{-}e^{-ikn}$ that propagate in both directions along the nanowire.  However, because of the symmetry breaking induced by the confinement, the wave vector $k$ is quantized according to $N$ values which can be real or complex, depending on the various parameters. A real solution for $k$ characterizes a stationary wave with a frequency inside 
the polaron band $\lambda=2\cos(k)$, i.e. $\omega_{0}-E_{B}-2\hat{J}_{N/2} < \omega < \omega_{0}-E_{B}+2\hat{J}_{N/2}$. A complex value 
of $k$ corresponds to a localized state which the wave function
is strongly localized on the side molecules and which the energy 
lies below or above the polaron band. A localized state is 
characterized by its localization length $\xi $, i.e. the inverse of the imaginary 
part of $k$, which defines the spatial extension of its amplitude. 

As detailed in Refs. \cite{kn:pouthierrg1,kn:pouthierrg2}, the occurrence of localized states corresponds to a critical transition between a stationary regime and a localized regime as the relevant parameters of the problem vary. Therefore, as shown in Appendix A, the Renormalization Group (RG) theory can be used in order to understand the transition and to predict the occurrence of localized states.

The conditions for the occurrence of high energy localized states is thus expressed as 
\begin{eqnarray}
&&\alpha+(1+\beta)^{2}-2=0 \nonumber \\
&&\alpha+(1+\beta)^{2}-2-\frac{2}{N-1}(1+\beta)^{2}=0 
\label{eq:PHASE1}
\end{eqnarray}
whereas the condition for the occurrence of low energy localized states is written as
\begin{eqnarray}
&&\alpha-(1+\beta)^{2}+2=0 \nonumber \\
&&\alpha-(1+\beta)^{2}+2+\frac{2}{N-1}(1+\beta)^{2}=0 
\label{eq:PHASE2}
\end{eqnarray}

As shown in Fig. 9, Eqs.(\ref{eq:PHASE1}) and (\ref{eq:PHASE2}) define the critical curves in the space of the parameters ($\alpha$, $\beta$) which separate phases exhibiting zero, one, two, three and four localized states. More precisely, as indicated in the figure, the full lines characterize the occurrence of high energy localized states located above the polaron band. The lines discriminate between three regimes connected to the presence of zero (0HE), one (1HE) or two (2HE) high energy localized states. By contrast, the dashed lines refer to the occurrence of low energy localized states, located below the polaron band, and discriminate between three regimes connected to the presence of zero (0LE), one (1LE) or two (2LE) low energy localized states. 

To explain the behavior of the phase diagram, let us consider the situation addressed in our model and which corresponds to $\alpha>0$ and $\beta>0$. The generalization of the entire parameter space is straightforward. When $\beta=0$, i.e. when the hopping constant is uniform in the nanowire, the lattice exhibits two defects corresponding to the side sites which the energy is blueshifted when compared to that of the core sites. Therefore, the phase diagram shows that two critical values for $\alpha$ discriminate between phases with zero, one and two high energy localized states.  
By contrast, when $\alpha=0$, the lattice exhibits two defects corresponding to a blueshift of the hopping constant which connects each side site to its nearest neighbor site. The phase diagram shows that two critical values for $\beta$ discriminate between phases with zero, two, and four localized states. In the two localized states regime, both a low energy and a high energy localized state occur whereas in the four localized states regime, the system exhibits two low energy and two high energy localized states. Finally, for both non vanishing $\alpha$ and $\beta$ values, intermediate situations take place. 

As detailed in Refs. \cite{kn:pouthierrg1,kn:pouthierrg2}, such a behavior can be interpreted
in terms of a semi-infinite lattice with defects at its side. For such a system, one or two localized states can occur depending on the values of the parameters. For instance, when $\beta=0$, one high energy localized state appears when $\alpha > 2$. In the same way, when $\alpha=0$, the lattice exhibits two localized states when $(\beta+1)^{2}>2$. One state is located below the polaron band (low energy localized state) whereas the second one appears above the polaron band (high energy localized state). 
The finite size nanowire can thus be viewed as the superimposition of two
semi-infinite lattice with defects at their finite sides. Thus, when the required relations are full filled, one (or two) localized state occurs on each side of the nanowire. However, states localized onto two different sides interfere inside the chain core. The strength of the coupling depends on the overlap of their respective amplitudes which leads to hybridization. 
For a strong overlap, the coupling can be
sufficient to strongly split the energies and push back half of the states into the
polaron band which thus loses their localized specificity. Therefore, the nanowire exhibits one (or two) localized state. By contrast, when the overlap
is weaker, the energies of the states remain outside the polaron band and
the nanowire supports two (or four) localized states. Since the occurrence of this latter regime depends on the overlap of the original localized states, the corresponding critical values depend on the system size (See Eqs.(\ref{eq:PHASE1}) and (\ref{eq:PHASE2})).
For instance, when $\beta=0$, it is straightforward to show that the semi-infinite lattice exhibits a high energy localized state when $\alpha>2$. Therefore, in the finite size nanowire, two original localized states occur on each side when $\alpha>2$. However, these two states interact due to their overlap inside the core of the lattice and their hybridization takes place which results in the formation of two states which appears as superimpositions of the two original localized states. As a result, when $2<\alpha<2+2/(N-1)$, the hybridization is strong enough so that the low energy state is pushed back into the polaron band. The lattice thus supports a single localized state, only. By contrast, when $\alpha>2+2/(N-1)$, the  hybridization is weaker and the two states remain above the polaron band.
 
Although the phase diagram displayed in Fig. 9 allows for a complete understanding of the localization transition, it involves parameters which are not independent. For instance, both $\alpha$ and $\beta$ depend on $E_{B}$ and on the temperature. However, as illustrated in Fig. 10 for $N=10$ (full line) and $N=20$ (dashed line), we can define a phase diagram in the ($E_{B}$, $T$) parameter space. In the parameter range considered here, the diagram exhibits two critical curves which discriminate between phases with zero, one and two localized states. 
The figures clearly shows that both the small polaron binding energy and the temperature favor the occurrence of localized states. However, these two parameters do not contribute in the same way to the localization mechanism. The influence of the small polaron binding energy in the localization processes is threefold. First, as when increasing $E_{B}$, the energy difference between the side and core sites increases. Then, strong $E_{B}$ values yield a strong dressing effect so that the core hopping constant is drastically reduced. Finally, as when increasing $E_{B}$, the inhomogeneous behavior of the polaron hopping constant is enhanced (see Fig. 2). All these features act in the same direction and favor the localization. By contrast, the influence of the temperature on the localization processes originates essentially in the dressing effect. Indeed, as when increasing the temperature the core hopping constant is reduced and the localization is enhanced. Note that the temperature favors a uniform behavior of the hopping constant and prevents the localization due to the $\beta$ parameter. 

In addition, the phase diagram illustrates the influence of the size of the nanowire. It shows that the lower critical curve depends slightly on $N$ whereas the upper critical curves exhibits a more important dependence. These two features have two distinct origins. Indeed, the size dependence of the lower critical curve originates in the dependence of the core hopping constant with respect to the lattice size. By contrast, the size dependence of the upper critical curve is essentially due to the overlap mechanism involved in the superimposition of the original states localized onto different sides of the nanowire. Therefore, Fig. 10 shows that this latter dependence decreases as the size of the lattice is increased, in agreement with Eqs.(\ref{eq:PHASE1}) and (\ref{eq:PHASE2}), so that the two critical curves tend to collapse. 

Finally, in the space of the parameters ($E_{B}$, $T$), Fig. 10 clearly shows that the finite size nanowire can exhibit a maximum of two localized states. This feature seems to be in contrast with the results displayed in the phase diagram connected to the general ($\alpha$, $\beta$) space where it is shown that a maximum of four localized states can be reached (see Fig. 9). In fact, since the parameters $\alpha$ and $\beta$ are expressed in terms of $E_{B}$ and $T$, the physical space ($E_{B}$, $T$) corresponds to a particular projection in the general space ($\alpha$,$\beta$). More precisely, the range for the variations of the physical parameters $E_{B}$ and $T$ generates the restriction $\alpha>0$, $\beta>0$ and $\alpha>>\beta$. In that case, Fig. 9 clearly shows that the nanowire supports a maximum of two high energy localized states.

To illustrate these features, let us estimate the order of magnitude of the key parameters for the CO/NaCl system from the knowledge of the vibron dynamics in the monolayer \cite{kn:pouthier4}. The CO molecules form a (2$\times$1) structure with a rectangular unit cell containing two molecules mutually perpendicular. Along the longer size of the cell, the vibron bandwidth is about 3 cm$^{-1}$ whereas it reaches 12 cm$^{-1}$ along the perpendicular direction. The frequencies of the external modes of the monolayer range from 36 cm$^{-1}$ for the translational motion parallel to the surface to 140 cm$^{-1}$ for the collective librational motion. The comparison with experimental spectra has shown that the phase relaxation is due to the coupling between the vibrons and the collective librons of the monolayer. Moreover, the dephasing constant exhibits a linear dependence vs the temperature in agreement with the deformation potential approximation used in the present work. Although the small polaron binding energy is unknown for this system, we expect a rather weak E$_{B}$ value, less than or about to 1 cm$^{-1}$, in agreement with the weak value of the dephasing constant equal to 0.2 cm$^{-1}$ at 55 K. As a consequence, we expect that the weak $E_{B}$ value prevents the formation of any localized state.

Note that the present model can also be applied to characterize the vibron-phonon dynamics in a finite size $\alpha$-helix protein (see for instance Ref. \cite{kn:davydov}). In that case, the high frequency modes correspond to either the C=O stretching vibrations (amide-I) or the N-H stretching modes of the peptide groups of the protein. The phonons characterize the collective dynamics of the external motions of the peptide groups. The parameters commonly used to described the system are J=7.8 cm$^{-1}$,  $\Omega_{c}=87-137$ cm$^{-1}$, E$_{B}$=15 cm$^{-1}$ for the C=O vibrons (see for instance Refs.\cite{kn:pouthier6,kn:pouthier7}) and J=5 cm$^{-1}$, $\Omega_{c}=100$ cm$^{-1}$, E$_{B}$= 84 cm$^{-1}$ for the N-H vibrons \cite{kn:hamm}. In these cases, we expect that the strong $E_{B}$ value allows for the formation of one or two localized states at biological temperature. Note that the adsorption of $\alpha$-helices or similar molecules could be considered.

\section{Conclusion} 

In the present paper, the small polaron theory was applied to describe the vibron dynamics in an adsorbed nanowire with a special emphasis onto finite size effects.
It has been shown that side molecules and core molecules do not experience the same interaction with the phonon bath which results in a different dressing effect. The redshift of the frequency of the core molecules is equal to the shift occurring in an infinite lattice whereas side molecules experience a frequency shift two times smaller. In addition, the inhomogeneous behavior of the polaron hopping constant has been established. The hopping constant between a side molecule and its nearest neighbor molecule is greater than the hopping constant which connects core molecules. Moreover, it has been shown that the core hopping constant depends on the size of the nanowire which indicates that the dressing mechanism is size dependent.
However, we have shown that the property of a lattice with translational invariance is recovered when the size of the nanowire is greater than a typical value $N^{*}\propto E_{B}k_{
B}T/\Omega_{c}^{2}$. These features have been attributed to the modification of the phonons induced by their confinement which is responsible for the occurrence of a finite number of stationary states. As decreasing the lattice size, the number of phonon modes decreases so that the vibron-phonon interaction is reduced when compared to that in a lattice with translational invariance which results in a softening of the dressing effect. 

The modification of the dressing mechanism was shown to be responsible for the occurrence of localized states. The nature and the number of localized states versus the relevant parameters of the problem have been summarized in a phase diagram. Although the lattice is able to supports a maximum of four localized states, it has been shown the for physical values of the parameters, the nanowire can exhibit zero, one or two high energy localized states. 

\appendix

\section{Decimation of the Schrodinger equation}

To predict the occurrence of localized states, a decimation \cite{kn:kadanoff,kn:stinchcombe,kn:pouthierrg1,kn:pouthierrg2} of the Schrodinger equation Eq. (\ref{eq:schrod2}) is performed. It consists in eliminating one site over two from the initial lattice to arrive at a scaled lattice that has twice the initial lattice spacing. In Eqs. (\ref{eq:schrod2}), we eliminate the amplitudes of the even sites by substituting their expressions in the Schrodinger equations for the odd sites and obtain a new set of equations for the odd sites only, as
\begin{eqnarray}
(\lambda^2-2) \psi_1 &=& (\alpha\lambda +(1+\beta)^2-2) \psi_1 +(1+\beta)\psi_{3}   \\
(\lambda^2-2) \psi_3 &=& (1+\beta) \psi_1 + \psi_5 \nonumber \\
(\lambda^2-2) \psi_5 &=& \psi_3 + \psi_7 \nonumber \\
...  &=& ... \nonumber \\ 
(\lambda^2-2) \psi_N &=&(\alpha\lambda +(1+\beta)^2-2) \psi_N +(1+\beta)\psi_{N-2} \nonumber 
\end{eqnarray}
Eq.(A1) characterizes the Schrodinger equation of a rescaled lattice with new parameters defining the RG transformation $R$ as
\begin{eqnarray}
\lambda^{(1)} &=& \lambda^2-2  \\
\alpha^{(1)}  &=& \alpha\lambda+(1+\beta)^2-2 \nonumber \\
\beta^{(1)} &=& \beta \nonumber  
\end{eqnarray} 
When the decimation is applied recursively $p$ times, with $N=2^{p+1}+1$, we obtain an ultimate scaled lattice formed by three sites, only, for which  
the eigenenergies are expressed as
\begin{eqnarray}
\lambda^{(p)} &=& \alpha^{(p)}   \\
\lambda^{(p)}&=& \frac{\alpha^{(p)}}{2}
\pm \sqrt{\frac{(\alpha^{(p)}}{2})^{2}+2(1+\beta^{(p)})^{2}} \nonumber  
\end{eqnarray}

At this step, a high energy localized state occurs when its energy is at least equal to the maximum of the polaron band. This condition is obtained when $\lambda =\lambda _c= 2$. At the critical point $\lambda =\lambda _c$, the scaled values of the
parameter $\lambda $ (Eq. (A2)) satisfy $\lambda^{(1)}=...=\lambda^{(p)}=\lambda _c$. 
This parameter $\lambda $ becomes scale invariant and is a fixed point $\lambda_{c}$ of the
RG transformation $R$ that verifies $\lambda_{c}=R(\lambda_{c})$. 
Indeed, no change in the critical parameters accompanies the 
length scaling so that the scaled lattice remains at a critical point.

From Eqs .(A3), the scaled values of the dynamical parameters at the critical 
point are expressed as
\begin{eqnarray}
\alpha^{(p)}  &=& 2-(1+\beta)^2-\frac{N-1}{2}(2-\alpha-(1+\beta)^2)  \\
\beta^{(p)} &=& \beta  \nonumber
\end{eqnarray}
Combining Eqs.(A3) and (A4) for $\lambda ^{(p)}=2$ leads
to the conditions for the occurrence of high energy localized states Eqs.(\ref{eq:PHASE1}). The condition (\ref{eq:PHASE2}) for the occurrence of low energy localized state can be obtained in a similar way.

\begin{figure*}[p]
\begin{center}
\includegraphics{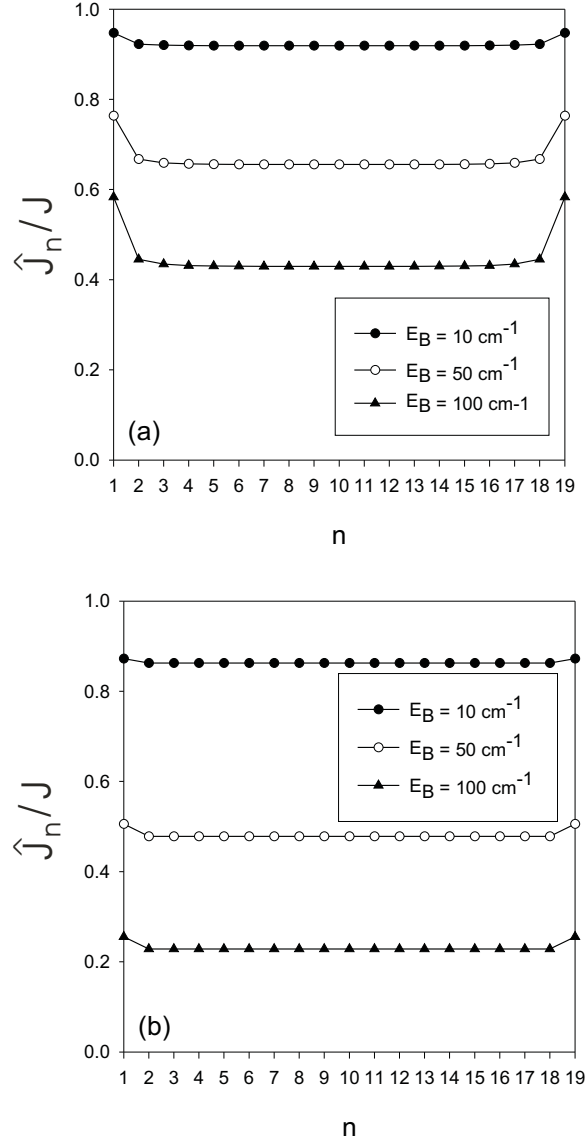}
\end{center}
\caption{Polaron hopping constant $\hat{J}_{n}$ versus the site position for $T=5$ K (a) and $T=50$ K (b) and for different values of the small polaron binding energy $E_{B}$. The number of molecules belonging to the nanowire is fixed to $N=20$. In all the figures, the undressed hopping constant is fixed to the typical value $J=4.0$ cm$^{-1}$ and the phonon cutoff frequency is set to $\Omega_{c}=100$ cm$^{-1}$.}
\end{figure*}

\begin{figure*}[p]
\begin{center}
\includegraphics{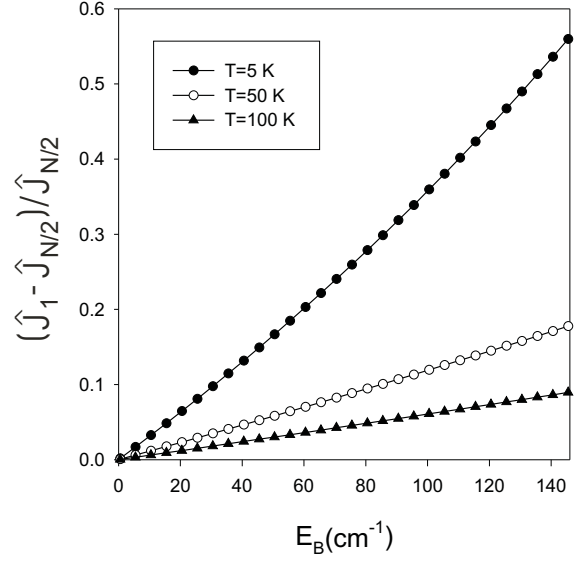}
\end{center}
\caption{Behavior of the difference between the side and the core hopping constants vs the small polaron binding energy $E_{B}$ for $N=30$ and $T=5$ K (full circles), $50$ K (open circles) and $100$ K (full triangles).}
\end{figure*}

\begin{figure*}[p]
\begin{center}
\includegraphics{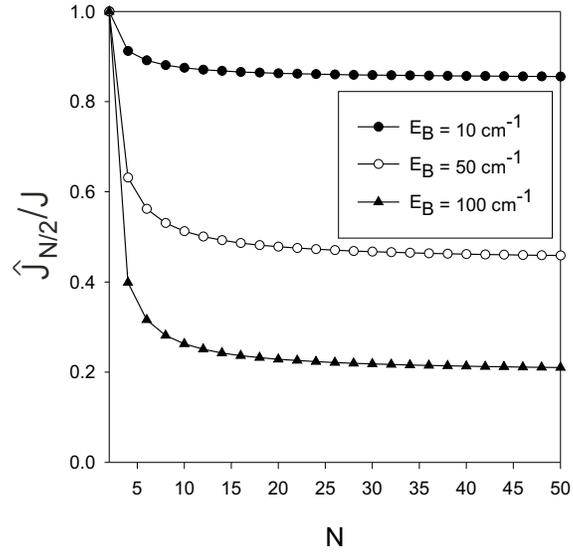}
\end{center}
\caption{Behavior of the core hopping constant vs the size $N$ of the nanowire size for $T=50$ K and for $E_{B}=10$ cm$^{-1}$ (full circles), $50$ cm$^{-1}$ (open circles) and $100$ cm$^{-1}$ (full triangles).}
\end{figure*}

\begin{figure*}[p]
\begin{center}
\includegraphics{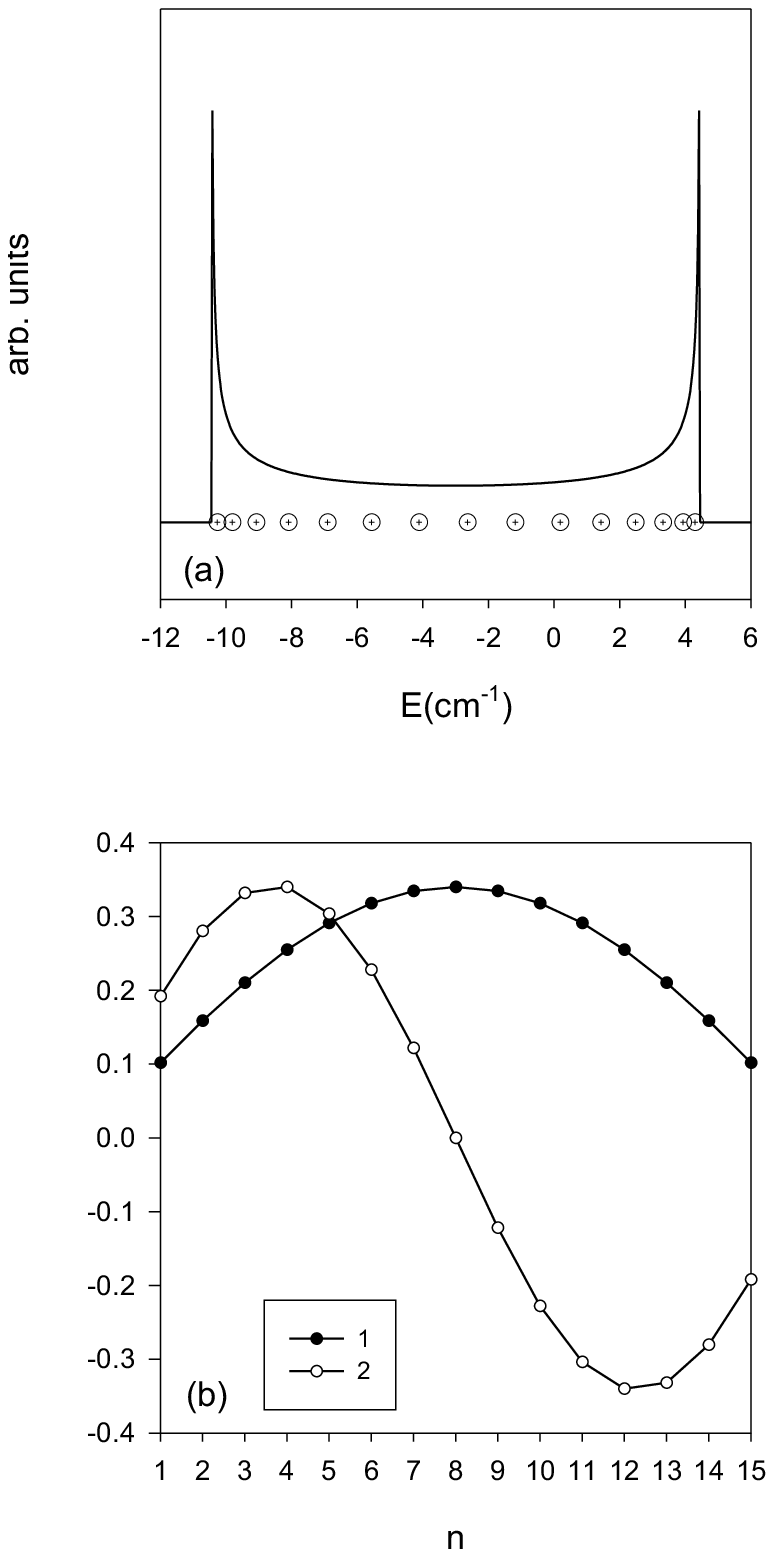}
\end{center}
\caption{(a) Small polaron eigenenergies for a nanowire containing $N=15$ molecules. The temperature is fixed to $T=100$ K and the small polaron binding energy is equal to $E_{B}=3$ cm$^{-1}$. The full line represents the density of states of the corresponding infinite lattice with translational invariance and characterized by the core hopping constant of the finite size nanowire. (b) Eigenfunctions of the two eigenstates with the higher energy.}
\end{figure*}

\begin{figure*}[p]
\begin{center}
\includegraphics{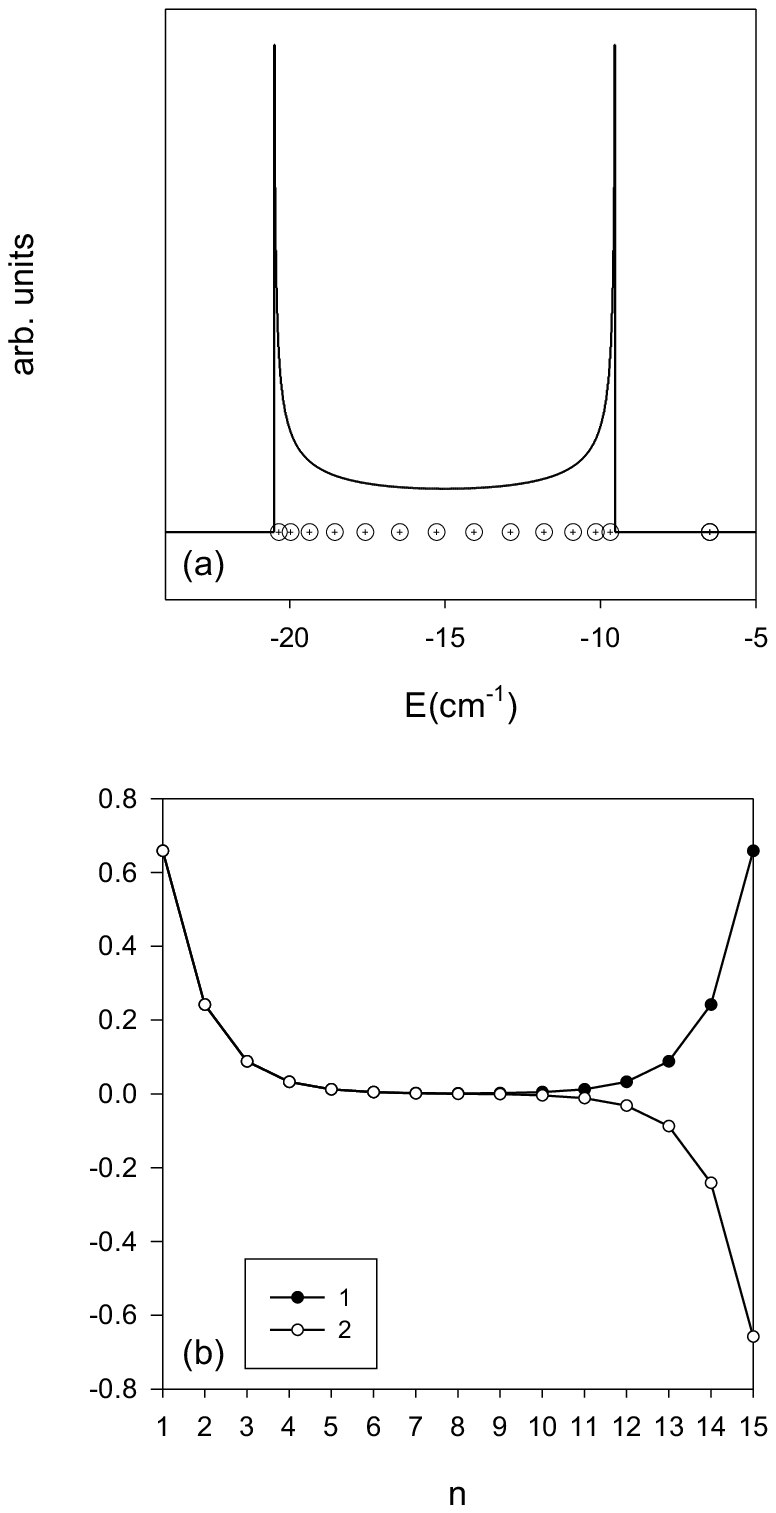}
\end{center}
\caption{(a) Small polaron eigenenergies for a nanowire containing $N=15$ molecules. The temperature is fixed to $T=100$ K and the small polaron binding energy is equal to $E_{B}=15$ cm$^{-1}$. The full line represents the density of states of the corresponding infinite lattice with translational invariance and characterized by the core hopping constant of the finite size nanowire. (b) Eigenfunctions of the two eigenstates with the higher energy.}
\end{figure*}

\begin{figure*}[p]
\begin{center}
\includegraphics{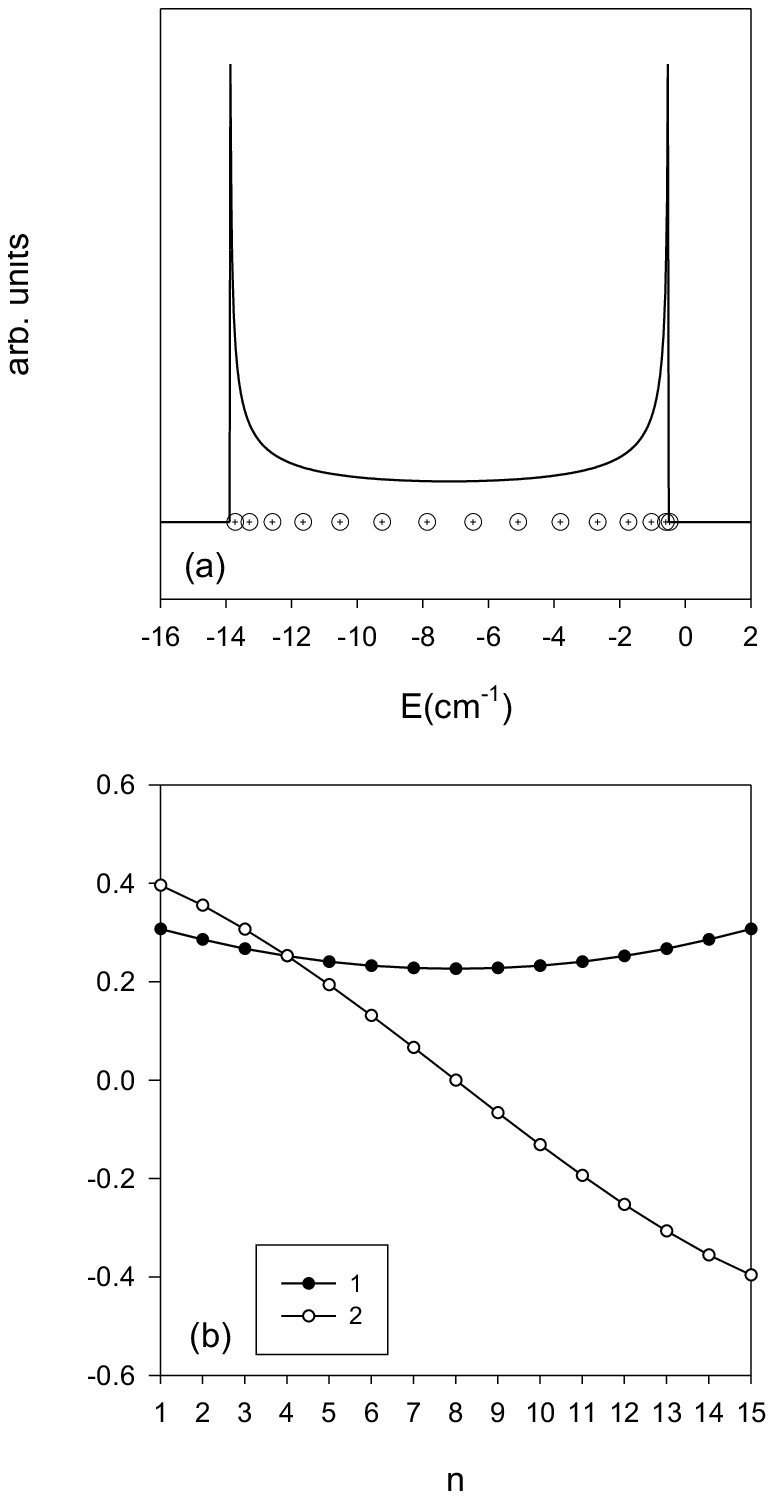}
\end{center}
\caption{(a) Small polaron eigenenergies for a nanowire containing $N=15$ molecules. The temperature is fixed to $T=100$ K and the small polaron binding energy is equal to $E_{B}=7.2$ cm$^{-1}$. The full line represents the density of states of the corresponding infinite lattice with translational invariance and characterized by the core hopping constant of the finite size nanowire. (b) Eigenfunctions of the two eigenstates with the higher energy.}
\end{figure*}

\begin{figure*}[p]
\begin{center}
\includegraphics{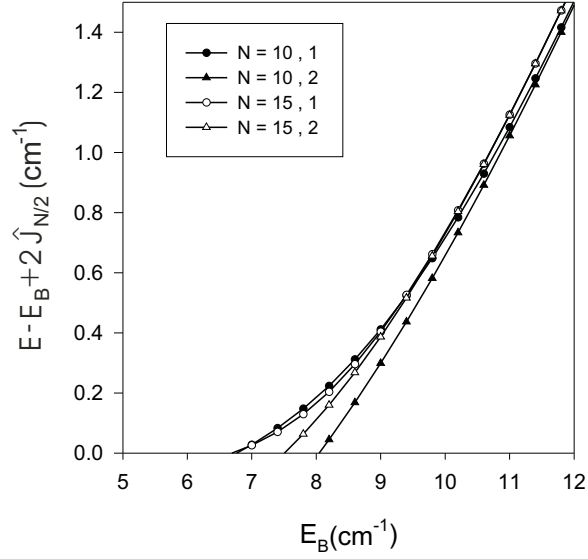}
\end{center}
\caption{Behavior of the energy of the localized states from the top of the polaron band vs the small polaron binding energy for two values of the lattice size $N=10$ (black symbols) and $N=15$ (white symbols). The temperature is fixed to $T=100$ K.}
\end{figure*}

\begin{figure*}[p]
\begin{center}
\includegraphics{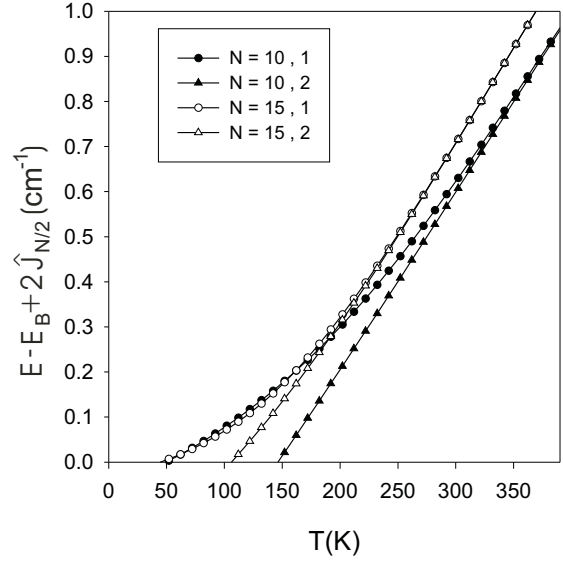}
\end{center}
\caption{Behavior of the energy of the localized states from the top of the polaron band vs the temperature for two values of the lattice size $N=10$ (black symbols) and $N=15$ (white symbols). The small polaron binding energy is fixed to $E_{B}=7$ cm$^{-1}$.}
\end{figure*}

\begin{figure*}[p]
\begin{center}
\includegraphics{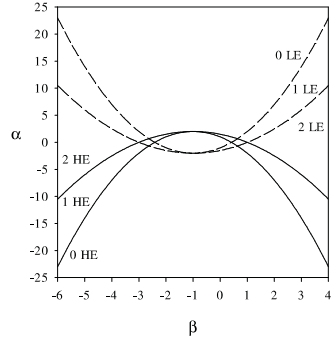}
\end{center}
\caption{Phase digram in the space of the parameters ($\alpha$, $\beta$) (see the text). The full lines characterize the occurrence of high energy localized states located above the polaron band. The lines discriminate between three regimes connected to the presence of zero (0HE), one (1HE) or two (2HE) high energy localized states. The dashed lines refer to the occurrence of low energy localized states, located below the polaron band, and discriminate between three regimes connected to the presence of zero (0LE), one (1LE) or two (2LE) low energy localized states. }
\end{figure*}

\begin{figure*}[p]
\begin{center}
\includegraphics{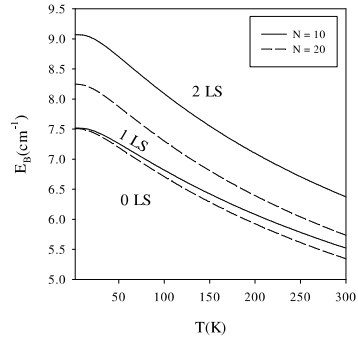}
\end{center}
\caption{Phase digram in the space of the parameters ($E_{B}$, $T$) for $N=10$ (full lines) and $N=20$ (dashed lines). For each $N$ value, the diagram exhibits two critical curves which discriminate between phases with zero, one and two high energy localized states.}
\end{figure*}

\end{document}